\documentclass{rsos}
\usepackage{amsmath}

\newcommand{\aanda}{\textit{Astron.\ Astrophys.\/}}
\newcommand{\aandas}{\textit{Astron.\ Astrophys.\ Suppl.\ Ser.\/}}
\newcommand{\aarev}{\textit{Astron.\ Astrophys.\ Rev.\/}}
\newcommand{\aj}{\textit{Astron.\ J.\/}}
\newcommand{\apj}{\textit{Astrophys.\ J.\/}}
\newcommand{\apjs}{\textit{Astrophys.\ J.\ Suppl.\ Ser.\/}}
\newcommand{\araa}{\textit{Ann.\ Rev.\ Astron.\ Astrophys.\/}}
\newcommand{\memras}{\textit{Mem.\ R.\ Astron.\ Soc.\/}}
\newcommand{\mnras}{\textit{Mon.\ Not.\ R.\ Astron.\ Soc.\/}}
\newcommand{\nature}{\textit{Nature\/}}
\newcommand{\newar}{\textit{New Astron.\ Rev.\/}}
\newcommand{\pasa}{\textit{Publ.\ Astron.\ Soc.\ Austr.\/}}
\newcommand{\pasp}{\textit{Publ.\ Astron.\ Soc.\ Pac.\/}}

\newcommand{\etal}{\textit{et al.\/}}

\begin{document}

\title{Extragalactic radio surveys in the pre-Square Kilometre Array era}

\author{
Chris Simpson$^{1}$}

\address{$^{1}$Gemini Observatory, Northern Operations Center, 670
  North A`\={o}h\={o}ku Place, Hilo, HI 96720-2700, USA}

\subject{Astronomy}

\keywords{surveys, radio continuum: galaxies, galaxies: active}

\corres{Chris Simpson\\
\email{csimpson@gemini.edu}}

\begin{abstract}
The era of the Square Kilometre Array is almost upon us, and
pathfinder telescopes are already in operation. This brief review
summarizes our current knowledge of extragalactic radio sources,
accumulated through six decades of continuum surveys at the
low-frequency end of the electromagnetic spectrum and the extensive
complementary observations at other wavelengths necessary to gain this
understanding. The relationships between radio survey data and surveys
at other wavelengths are discussed. Some of the outstanding questions
are identified and prospects over the next few years are outlined.
\end{abstract}


\begin{fmtext}
\section{Introduction}

Radio surveys provide a unique window to the distant Universe: the
extremely strong evolution of the radio luminosity function (RLF)
means that even shallow surveys contain distant objects. This was
recognized in the early days of radio surveys, when the brightest
extragalactic radio source, Cygnus~A, was identified with a galaxy in
a cluster at $z=0.056$ \cite{baademinkowski54}, making it one of the
most distant objects known at the time. In comparison to galaxies in
other distant clusters, the redshift of Cygnus~A was very easily
measured from its rich emission line spectrum. Although the invention
of the photometric redshift method \cite{baum57} allowed estimates of
distances to normal galaxies too faint for spectroscopy, radio
galaxies frequently displayed bright emission lines that provided
easy-to-measure and accurate spectroscopic redshifts. When the
redshift of 3C~295 was measured at $z=0.4614$ \cite{minkowski60} this
marked the beginning of a quarter of a century when a radio source
marked the boundary of the known Universe.

\end{fmtext}


\maketitle

A catalogued continuum radio source by itself only has position and
intensity information and is of limited scientific value without an
optical counterpart.  With the error ellipses from early surveys
typically several arcminutes in size, it was difficult to pinpoint the
origin of the radio emission unless it was obviously associated with a
bright galaxy or cluster of galaxies. Work therefore focused on the
radio spectral energy distributions of sources, finding that most
possessed a steep spectrum best explained by synchrotron emission, and
analysis of the source counts showed strong evolution inconsistent
with a Euclidean, Steady State Universe \cite{scheuer57}. The
construction of long-baseline interferometers allowed for more
accurate positions and the ability to determine counterparts to a
larger fraction of catalogued sources.

Of the many radio surveys undertaken during the late 1950s and 1960s,
the 3CR survey \cite{edge3c,bennett3c}, covering the northern sky to a
bright 178-MHz flux density limit $S_{178} > 9$\,Jy, has been perhaps
the most enduring. Sources from surveys at fainter flux limits are
either intrinsically less luminous (resulting in weaker emission lines
and more difficult redshift determinations) or more distant (pushing
the strongest emission lines beyond the limited range of optical
spectrographs), and the shape of the source counts results in a rapid
increase in source density on the sky with decreasing flux.  The
catalogue's modest size also presented the possibility of high
redshift completeness, although the final spectroscopic redshift in
the more rigorously defined 3CRR catalogue \cite{lrl83} was not
obtained until 1996 \cite{sr_4c1366} and another was corrected a few
years later \cite{willott_3c318}. And, despite its size, the strong
evolution of the radio source population means that this catalogue
contains a significant number of high-redshift objects -- over
one-fifth of its sources (37/173) have $z > 1$, with one (3C~9) at $z
> 2$. This fraction of distant sources does not exist in a
flux-limited catalogue at any other wavelength, which explains the
reason for the use of radio surveys to search for distant objects.

Even in those early days, the number of catalogued radio sources far
outstripped the ability to make robust optical identifications or
perform spectroscopy of them. The specific follow-up of other surveys
has depended on the scientific aims of the research groups
involved. Some have sought to study the evolution of radio sources as
a class and so require high redshift completeness, necessitating the
study of selected areas of sky covered by the deeper radio
surveys. Others have focused on identifying interesting objects (most
notably very distant sources) and have developed ways to
preferentially select such sources from the radio surveys, leaving the
vast majority unstudied. More recently, groups have been able to use
data from large public surveys to provide the necessary complementary
data and, in the last few years, very deep radio observations have
been undertaken in fields chosen because of their excellent
multiwavelength data, specifically to benefit from these complementary
imaging and spectroscopic observations.

The purpose of this review is not to comprehensively list every radio
survey that has ever been undertaken. Instead it aims to summarize our
current level of understanding regarding the multiple source
populations that are detected in radio surveys while also providing
historical context.

\section{Powerful radio sources}

Sub-arcminute imaging of bright extended extragalactic radio sources
at low radio frequencies reveals them to broadly fall into two
morphological classes. Those with the brightest regions near the
centre of the source (`edge-darkened') were named `Class~I' sources by
Fanaroff \& Riley, while those with the brightest regions near the
furthest extent of the radio emission (`edge-brightened') were called
`Class~II' \cite{fanaroffriley74}. These have subsequently become
known as FR\,I and FR\,II morphologies, after the authors of the study
who found a strong correlation between morphological class and
luminosity, with FR\,II sources being the most luminous. The boundary
between the two classes is not sharply-defined but there are roughly
equal numbers of sources at a 178-MHz luminosity of $L_{178} \sim
10^{26}$\,W\,Hz$^{-1}$.

A second dichotomy among radio sources is also seen in their optical
spectra. Sources either display rich emission-line spectra similar to
the narrow-line spectra of Seyfert galaxies and quasars, or are almost
entire devoid of emission lines, possibly only displaying weak
[O{\sc~ii}]~$\lambda$3727 emission.  Hine \& Longair
\cite{hinelongair79} named these `Class~A' and `Class~B' sources,
respectively, and again radio luminosity is a key factor, with the
most powerful sources belonging to Class A and the transition from
Class~B-dominated to Class~A-dominated gradually taking place in the
decade of luminosity above the Fanaroff--Riley break. This
nomenclature never caught on and, when a quantitative classification
method was first considered \cite{laing94}, the terms
`high-excitation' and `low-excitation' were used to describe the
spectra. `HE(R)G' and `LE(R)G', for high- and low-excitation (radio)
galaxy, respectively, are now the most commonly-used names in the
literature. The similarity between the radio luminosities at which the
transitions between Hine \& Longair and Fanraoff \& Riley classes
occur is believed to be a coincidence but has led to frequent and
erroneous use of the terms interchangeably. The FR class is also
affected by extrinsic factors such as the circumgalactic density field
\cite{ledlowowen96,kaiseralexander97}, while the HERG/LERG
classification depends on the availability of ionizing photons and
hence the properties of the system only on parsec scales.

While early radio surveys were generally conducted at fairly low
frequencies ($\sim 100$\,MHz), samples were also constructed at higher
frequencies, most notably the 2.7-GHz Parkes catalogue \cite{ekers69}.
Surveys at different radio frequencies can produce significant
differences in the samples selected. A powerful radio source consists
of extended, steep-spectrum emission (typically $\alpha \approx 0.7$,
where $S_\nu \propto \nu^{-\alpha}$) and a jet-producing core with a
much flatter spectrum ($\alpha \approx 0$). The relative core-to-lobe
flux ratio, $R$, therefore increases with increasing frequency.
Although the intrinsic luminosity of the core is less than 1\,per cent
of the total luminosity, even at a fairly high frequency like 5\,GHz
\cite{morganti97,cjs00}, its apparent luminosity can be dramatically
increased by Doppler boosting, up to a factor of $\sim 2\gamma^4$,
where $\gamma \sim 5$ is the Lorentz factor of the
synchrotron-emitting particles in the jet
\cite{scheuerreadhead79,orrbrowne82}.  A highly favourable orientation
is required to boost the core emission by a sufficiently large factor
to make its observed flux comparable to the extended emission, but the
source counts are steep enough that bright high-frequency-selected
samples have a significant fraction of sources with core-dominated
morphologies. For example, the 178-MHz-selected 3CRR \cite{lrl83} and
2.7-GHz-selected Peacock \& Wall \cite{peacockwall81} samples have
similar source densities on the sky, but the fractions of sources
whose morphology is dominated by an unresolved core at GHz frequencies
are 11\,per cent (19/173) and 49\,per cent (83/168), respectively,
while the fractions of flat spectrum ($\alpha < 0.5$) core-dominated
sources are 4\,per cent (7/173) and 27\,per cent (46/168). Although
there are only two sources in the 3CRR catalogue for which Doppler
boosting of their cores has raised their total fluxes above the flux
limit (3C~345 and 3C~454.3), the fraction is significantly higher in
samples selected at GHz frequencies. High-frequency-selected samples
are therefore not as clean as lower-frequency ones in providing the
observed radio luminosity as a proxy for the intrinsic power of the
central engine.


\begin{figure}
  \centering\includegraphics[width=0.9\textwidth]{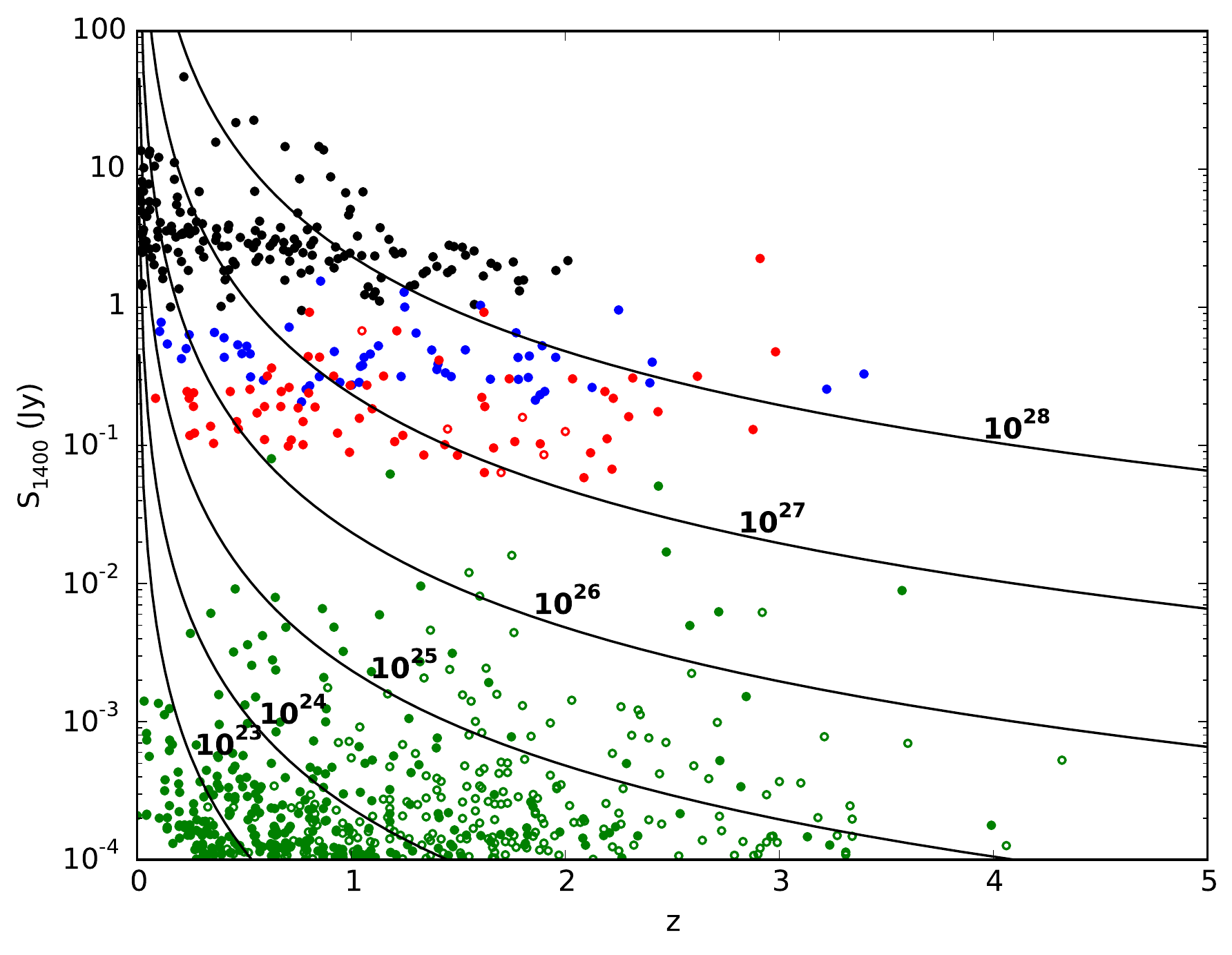}
  \caption[lz]{Observed 1400-MHz flux density against redshift for
    radio sources from four surveys: 3CRR \cite{lrl83} (black), 6CE
    \cite{sr_6ce} (blue), the 7C Redshift Survey \cite{willott01}
    (red), and the Subaru/\textit{XMM-Newton\/} Deep Survey
    \cite{cjs06} (green). While the SXDS survey has a flux limit at
    1400\,MHz, the other three surveys were selected and flux-limited
    at much lower frequencies. The existence of a correlation between
    radio spectral index and redshift results in the apparent 1400-MHz
    flux limits of these lower-frequency-selected samples decreasing
    with redshift, but it is impossible to determine from a single
    flux-limited sample whether the correlation is between spectral
    index and redshift or luminosity. Filled symbols indicate
    spectroscopic redshifts, while open symbols represent sources with
    redshifts estimated from broad-band photometry. Lines of constant
    1400-MHz luminosity are shown in steps of one dex from $10^{23}
    \leq L_{1400} \leq 10^{28}$\,W\,Hz$^{-1}$, as indicated (a
    spectral index $alpha = 0.7$ and a flat $\Lambda$CDM cosmology
    with $H_0 = 70$\,km\,s$^{-1}$\,Mpc$^{-1}$ and $\Omega_{\rm m} =
    0.3$ are assumed).}
  \label{fig:lz}
\end{figure}

Even in very early studies with incomplete data there was evidence
that the most luminous radio sources displayed a much stronger cosmic
evolution than did lower-luminosity sources \cite{longair66}, and this
result grew in strength as the data improved
\cite{lrl83,dunloppeacock90,clewleyjarvis04}. However, the steepness
of the radio source counts at bright fluxes results in a very strong
correlation between redshift and luminosity in any flux-limited sample
since a large fraction of the sample is within a factor of two of the
flux limit (three-quarters in the case of 3CRR, but only one-half for
the much fainter SXDS sample) (Fig.~\ref{fig:lz}). In any single,
reasonably-sized flux-limited sample of a few hundred radio sources,
it is therefore impossible to disentangle
\textit{redshift\/}-dependent effects from
\textit{luminosity\/}-dependent effects since the correlation between
luminosity and redshift swamps all other relationships
\cite{kmb99}. The use of multiple flux-limited surveys in a `wedding
cake' pattern allowed more complete sampling of the $L$--$z$ plane and
better observational constraints on the evolution of the radio
luminosity function. Advances in our understanding of the unification
of active galactic nuclei led to the positing of dual-population
models, with HERGs and LERGs being treated separately and undergoing
very different cosmic evolution \cite{willott01}.

\section{Deep surveys and the faint radio source population}

\begin{figure}
  \centering\includegraphics[width=0.9\textwidth]{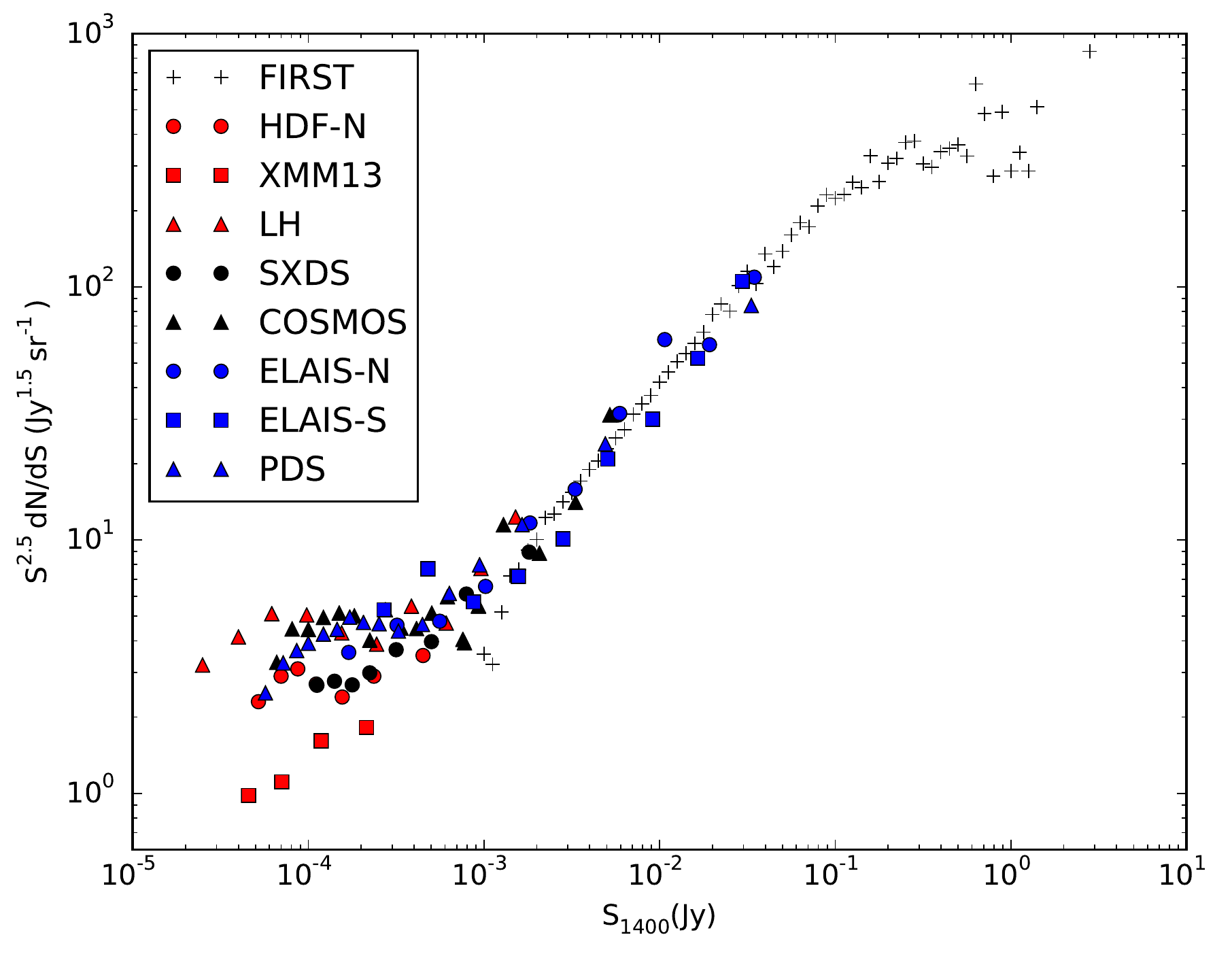}
  \caption[sourcecounts]{Euclidian-normalized source counts from
    selected 1.4-GHz radio surveys. To keep the figure clean, error
    bars are not plotted but only points where the uncertainty,
    including sampling variance \cite{dezotti10}, is less
  than one-third of the measured counts are plotted. Plus symbols
  indicate counts from the Faint Images of the Radio Sky at Twenty
  Centimeters Survey (FIRST; \cite{first}), while the other points
  have been colour-coded by survey area. Red points indicate counts
  from surveys covering less than $\sim1$\,deg$^2$ of the Hubble Deep
  Field North \cite{richards00}, the \textit{XMM-Newton\/} 13-hour
  field \cite{seymour08}, and the Lockman Hole \cite{ibar09}. Black
  points indicate surveys of $\sim1$\,deg$^2$ from the
  Subaru/\textit{XMM-Newton\/} Deep Survey \cite{cjs06} and the
  VLA-COSMOS survey \cite{bondi08}. Blue points indicate surveys of
  several square degrees, including the European Large Area
  \textit{ISO\/} Survey Northern \cite{ciliegi99} and Southern
  \cite{gruppioni99elais} regions, and the Phoenix Deep Survey
  \cite{hopkins03}. Note that the scatter at the faintest radio fluxes
  is much larger than predicted by cosmic variance \cite{heywood13}.}
  \label{fig:sourcecounts}
\end{figure}

The Leiden--Berkeley Deep Survey (LBDS) \cite{windhorst84} consisted
of nine pointings of the Westerbork Synthesis Radio Telescope (WSRT),
in four regions of the sky with deep multicolour photographic plates,
reaching sub-millijansky levels at a frequency of 1.4\,GHz.
Conducting the survey at this frequency provided not just the best
sensitivity, but also resulted in more accurate positions, permitting
reliable optical identifications. This became the frequency of choice
for deep radio maps and the adjectives `microjansky' or
`sub-millijansky' are normally assumed to refer to the flux densities
of radio sources around this frequency, with 1\,mJy being the flux
density at which the source counts flatten (when plotted in a
Euclidian-normalized manner, as in Fig.~\ref{fig:sourcecounts}),
indicating the presence of a new population. With the aid of VLA data,
the LBDS probed well into this regime and the multicolour imaging
revealed a change in the population at a flux density of a few mJy
\cite{windhorst85,kron85}.

The counterparts to bright radio sources were dominated by red
galaxies that followed the well-known Hubble relation for massive
elliptical galaxies \cite{sandage72iii}. However, at the fainter flux
densities where the source counts began to flatten, blue objects
became an increasingly important contributor. These had lower radio
luminosities than the red galaxies, and the more distant sources
showed peculiar optical morphologies. Many objects, however, remained
unidentified in the optical plates, limiting the general conclusions
that could be drawn.

Attempts to further understand the nature of faint radio sources were
hampered by the limitations of complementary data and available
instrumentation. A notable effort \cite{benn93,mrr93} demonstrated the
existence of a fairly heterogeneous mix of objects at $S_{1400}
\lesssim 1$\,mJy, although their reliance on optical counterparts from
digitized photographic plates eliminated nearly 80\,per cent of their
radio source sample from identification. They nevertheless concluded
that spiral galaxies (predominantly star-forming objects, but
including Seyferts) dominate the faint radio source counts. Linking
these directly to the objects responsible for the \textit{IRAS\/}
60-$\mu$m counts, they surmised that the strong evolution displayed is
driven by galaxy interactions and mergers. A somewhat contrary
conclusion was derived from a deeper (in both radio and optical
limiting fluxes) radio survey of the Marano field, which concluded
that the unidentified radio sources were probably distant elliptical
galaxies \cite{gruppioni99}.

Deeper optical imaging, including with the \textit{Hubble Space
  Telescope\/}, provided much higher identification rates and
detections at higher redshifts
\cite{windhorst95,richards98,richards99}. The morphological
information granted by the \textit{HST\/} images showed the hosts of
the very faintest radio sources to typically be disks, often
displaying signs of recent star formation. However, these studies were
confined to very small areas of the sky, usually a few arcminutes
across, and therefore most sources had flux densities $S_{1400} \ll
1$\,mJy. This is well below the source counts' change in slope and
fainter than the sources analyzed in the earlier studies, leading to
continued uncertainty as to the make-up of the population.

Another reason for the lack of progress towards a definitive
understanding of the faint radio source population during the 1990s
was sociological. For decades, the only spectroscopically-confirmed
high-redshift objects were radio sources and radio-quiet quasars, but
the discovery of large numbers of distant galaxies via the Lyman break
technique \cite{steidel96} changed this. Since the tight correlation
between black hole and galaxy masses
\cite{magorrian98,ferrarese00,gebhardt00} had not yet been discovered,
the black holes in active galaxies were viewed as a nuisance rather
than a fundamental component of galaxy evolution. The study of
extragalactic radio sources as a class quickly fell from favour and it
became difficult to obtain telescope time for follow-up studies.

Progress was made by using existing or planned deep, wide-area
extragalactic surveys to leverage new radio data. The prime-focus
camera on Subaru Telescope, Suprime-Cam, was a key ingredient in these
surveys as it provided an unprecedented combination of depth and
area. From a radio perspective, these optical images could provide a
high identification rate for statistically significant samples of
radio sources at flux densities spanning the flattening in the source
counts, and were therefore enormously helpful. Two fields provided the
best ancillary data for studies of the faint radio source population:
the Subaru/\textit{XMM-Newton\/} Deep Survey (SXDS \cite{furusawa08})
and the Cosmic Evolution Survey (COSMOS \cite{taniguchi07}). Covering
more than one square degree each, they allowed identification of a
sufficiently large sample of radio sources to quantify the composition
of the microjansky population.

The radio survey of the SXDS \cite{cjs06} was undertaken with the VLA
and comprised only 60\,hours of radio data, yet at the time it was the
deepest degree-scale radio survey, and therefore the first to contain
a meaningful number of radio sources with fluxes below the flattening
in the source counts. The deep multicolour imaging provided reliable
optical counterparts for all but a handful of sources and a simple
visual inspection revealed a significant minority population of blue,
star-like objects that were obviously optically-luminous quasars.
However, unlike the powerful radio-loud quasars found in shallow radio
surveys whose radio luminosities are typically above the FR break,
these objects were much less luminous, with ratios of radio to optical
luminosity that put them clearly within the radio-quiet regime (see
\S\ref{sec:rlrq}). The contribution to the radio source counts from
this population had not previously been well determined; indeed, it
had often been neglected with models that did not include this
population successfully fitting the observed counts \cite{seymour04}.
On the other hand, estimates based on the observed X-ray source counts
implied that it could be significant over a small flux range just
below the flattening in the source counts, possibly dominant depending
on the number of Compton-thick sources (which would contribute to the
radio source counts but not the X-ray counts) \cite{jarvisrawlings04}.
The SXDS observations suggested that at least 20\,per cent of the
sources just below the flattening of the source counts were luminous
active galactic nuclei (AGN) that lacked powerful radio jets, and this
estimate was subsequently supported by studies in the COSMOS field
\cite{smolcic08,smolcic17}. Although they do not dominate the source
counts, this population follows the strong strong cosmic evolution
traced by optically-selected quasars and can therefore make a
significant contribution to the RLF at redshifts $z \sim 2$.

\section{Star-forming galaxies}

Locally, radio emission is often observed to be associated with star
formation, with synchrotron emission from electrons accelerated in
supernova-driven shocks being the dominant component. The radio
spectrum of a star-forming galaxy therefore has a similarly steep
spectral index as an AGN, and both also flatten at higher frequencies
due to the increased importance of a flat-spectrum component: the
jet-producing core in an AGN, and thermal bremsstrahlung emission in a
star-forming galaxy. Since the cosmic star-formation rate density is
known to increase rapidly with redshift \cite{hopkinsbeacom06}, the
radio sky should be full of star-forming galaxies.

As the pioneering work in the Hubble Deep Field showed
\cite{richards99}, the dominant sources at the flux levels reached by
the deepest surveys ($S_{1400} \sim 10\,\mu$Jy) are star-forming
galaxies spanning a range of redshifts. For galaxies at $z > 2$ to be
above the flux threshold of these surveys, they must possess
star-formation rates (SFRs) of several hundred M$_\odot$\,yr$^{-1}$,
and this population therefore overlaps significantly with that of the
submillimetre galaxies (SMGs). The tight correlation between the radio
and far-infrared luminosities of star-forming galaxies \cite{helou85}
coupled with the vastly different $k$-corrections in the two
wavelength regimes allows the ratio of radio-to-submillimetre flux
densities to be used as a crude redshift indicator \cite{carilli_yun},
with fainter radio sources (for a given submillimetre flux) likely to
be more distant. Since single-dish submillimetre detections have large
astrometric uncertainties, the interferometric radio observations have
been essential to localize the optical/near-infrared counterpart for
additional analysis, including redshift determination. While some
groups used the VLA to make complementary observations of regions that
had already been surveyed at submillimetre wavelengths
\cite{smail00,ivison8mjy}, others identified distant star-forming
galaxies directly from their radio emission and optical faintness
\cite{barger00,chapman01}.

These radio detections were essential in permitting spectroscopic
observations to determine redshifts and provide statistical
information about the SMG population \cite{chapman03}, although one
thing that the high-resolution radio surveys failed to reveal was the
high level of multiplicity among single-dish submillimetre sources
($\sim40$\,per cent; \cite{hodge13}). The scatter in the
far-infrared--radio correlation coupled with the modest
signal-to-noise ratios of the radio detections makes it unlikely that
more than one component would be detected if the single-dish detection
was a blend of multiple sources. Nevertheless, the large primary beam
of a radio antenna makes deep radio surveys an excellent way to
produce large samples of star-forming galaxies at moderate-to-high
redshift. Existing radio surveys are able to derive the luminosity
function of star-forming galaxies out to $z \sim 5$ \cite{novak17}
(although the data do not probe below $L^*$ at $z>2$) and these
objects will dominate the next generation of radio surveys.

\section{Evolution of the radio luminosity function}

\begin{figure}
  \centering\includegraphics[width=\textwidth]{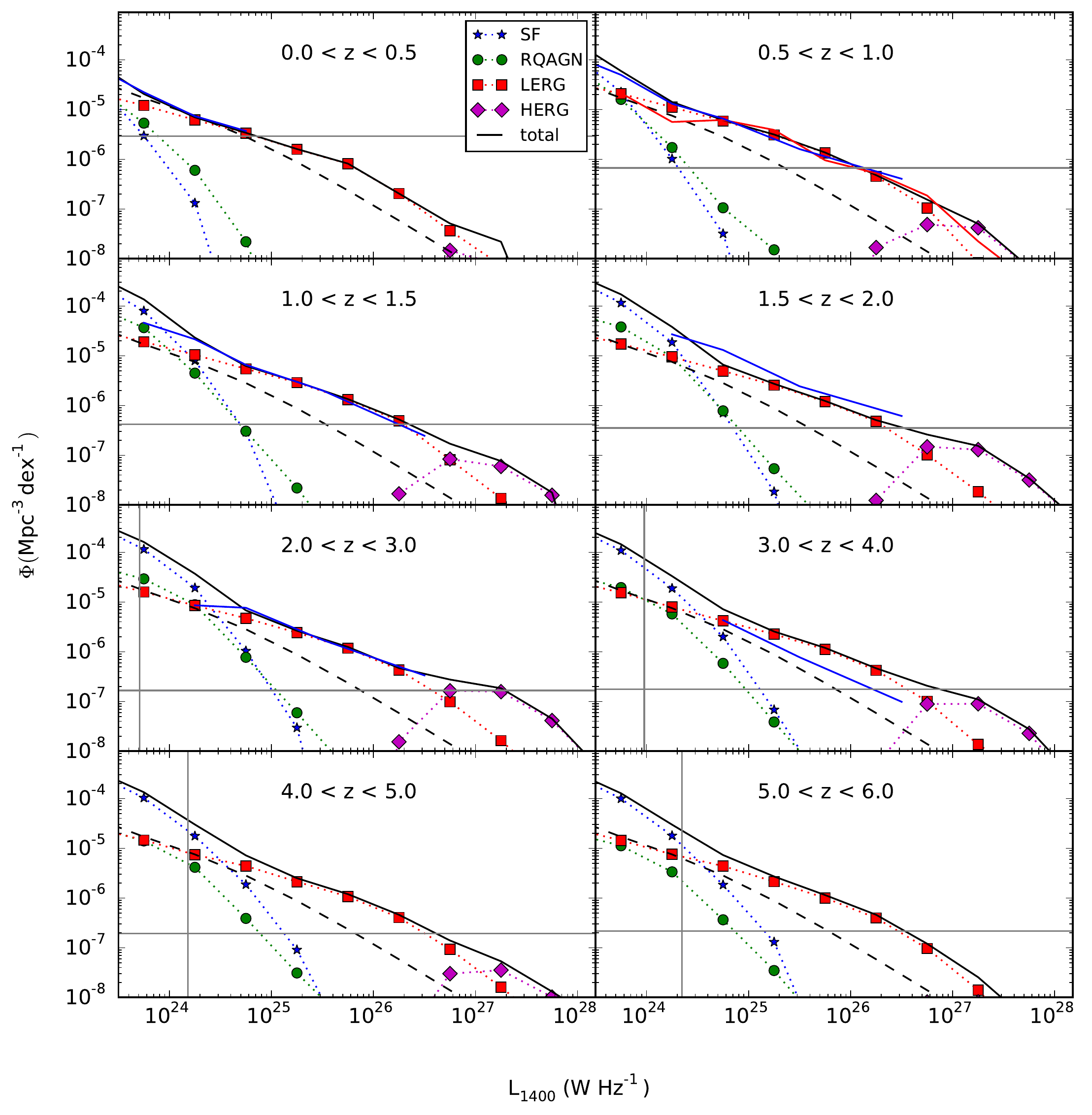}
  \caption[lfevol]{The 1.4-GHz radio luminosity function in different
    redshift ranges, as derived from the SKA Simulated Skies models
    \cite{wilman08}. The symbols denote different types of radio
    source as indicated by the key in the top-left panel, while the
    black solid line shows the overall luminosity function, and the
    black dashed line represents the parametrized redshift zero
    luminosity function from the 6-degree Field Galaxy Survey
    \cite{mauchsadler07}. The red line in the $0.5<z<1.0$ panel shows
    the observed radio luminosity function in this range
    \cite{best14}, while the blue line in each of the first six panels
    shows the luminosity function in the Subaru/\textit{XMM-Newton\/}
    Deep Field \cite{cjs12}. In each panel, the grey vertical line
    shows the luminosity of a source with $S_{1400}=10\,\mu$Jy at the
    distant edge of the redshift bin (if absent, this limit is off the
    left-hand edge of the panel), while the grey horizontal line shows
    the space density corresponding to one source per square degree
    per luminosity bin.}
  \label{fig:lfevol}
\end{figure}

Understanding the composition of radio surveys allows predictions for
how the RLF will evolve, and what the radio source counts will be at
currently unattainable flux levels, by using observations at other
wavelengths. Such knowledge is important for designing surveys for
future facilities like the SKA. Star formation in the Universe can be
traced by infrared and ultraviolet continuum radiation and emission
lines such as H$\alpha$ and [O{\sc~ii}], and accurate measurements
exist to $z>2$ (a recent review is given in
ref.~\cite{madaudickinson14}). For radio-quiet AGN, the evolution of
the optical quasar luminosity function has been well-determined to
$z>5$, and much telescope time has been awarded to X-ray surveys and
spectroscopic follow-up to attempt to quantify the obscured AGN
population. However, since X-ray emission can be affected by dust
and/or gas absorption while radio emission is not, the uncertainty
regarding the optically-obscured and Compton-thick AGN fractions
affects the predicted radio counts. The evolution of the most powerful
(FR\,II) radio sources has been determined out to $z \sim 3$ from
follow-up of radio surveys and, while uncertainty still remains over
whether a `redshift cut-off' exists, the effect of this uncertainty on
the source counts is negligible.

The SKA Simulated Skies (S$^3$ \cite{wilman08}) have used these
observations to produce a model of the radio sky over a wide range of
frequencies.  At 1.4\,GHz the model extends down to 10\,nJy, more than
three orders of magnitude fainter than the deepest surveys currently
in existence. At other frequencies, they surpass current data by even
greater amounts. The prediction of this model for the evolution of the
1.4-GHz luminosity function is shown in Fig.~\ref{fig:lfevol}, with
some current observational measurements and limitations shown.

Significant progress has been made in measuring the evolution of the
RLF by combining wide-area radio surveys with large spectroscopic
datasets that provide redshifts and source classifications. In the
very local Universe, the combination of the NRAO VLA Sky Survey (NVSS;
\cite{condon_nvss}) with the 6-degree Field Galaxy Redshift Survey
(6dFGRS) over 7000\,deg$^2$ has produced separate measurements of the
RLF for AGNs and star-forming galaxies with $L_{1400} >
10^{20}$\,W\,Hz$^{-1}$ \cite{mauchsadler07}. More sensitive
spectroscopic surveys have provided redshifts and classifications for
sources out to $z \sim 0.8$ \cite{pracy16} which, although considered
`modest', represents a look-back time equal to half the age of the
Universe. At $z > 1$, however, it becomes increasingly difficult to
determine redshifts and, particularly, classifications, since
diagnostic features such as the [O{\sc~ii}] and [O{\sc~iii}] emission
lines move into the near-infrared.

Over much of the luminosity range shown the LERG population dominates,
although the cosmic evolution of these sources has been the focus of
fairly limited study because, unlike the other classes of source, they
do not possess the strong emission lines that make redshift
determination simple. Deep-field studies using photometric redshifts
have suggested a decline in the space density of these sources
beyond $z\sim1$ \cite{cjs12,rigby11}, although work with higher
spectroscopic completeness redshifts indicates this population
experiences luminosity-dependent evolution, with the turnover taking
place at $z\sim0.7$ for sources below the characteristic luminosity of
$L_{1400} \approx 10^{26}$\,W\,Hz$^{-1}$ \cite{best14,pracy16}. These
sources are interesting because they display the clearest evidence of
AGN feedback as the radio-emitting plasma expands and creates
`bubbles' within the surrounding intergalactic medium
\cite{bohringer93,mcnamara00}. It has been suggested that this form of
AGN activity might be episodic in nature and, if so, the time-averaged
heating rate has been shown to balance the cooling of gas from the hot
halo, creating a self-regulating feedback loop that limits the stellar
mass of galaxies \cite{best05,best06}.

Robust observational measurements of the RLF at high redshifts and
modest luminosities are limited by the availability of complementary
data at other wavelengths and, in particular, spectroscopy. Even when
a spectrum is available, it often fails to provide a redshift, let
alone a classification. Unlike the powerful FR\,II HERGs that dominate
the bright radio surveys at these redshifts, the microjansky sources
usually have weaker emission lines, and fewer of them, because
star-formation processes do not produce the rich spectrum of an AGN,
and the emission-line producing region is often completely obscured by
dust \cite{ams06}. Photometric redshifts are therefore an essential
part of any study but all studies have a few per cent of `catastrophic
outliers' among the objects with spectroscopic redshifts and, of
course, these are not an unbiased subset so the true fraction of such
sources is unknown. While increasing the sample size and/or using the
full posterior $P(z)$ distribution can mitigate the effect of random
photometric redshift errors on the determination of luminosity
functions, these do not alleviate the errors introduced by outliers.
Apart from the primary effects of assigning an incorrect redshift, and
therefore luminosity, to a particular source, there can be secondary
effects. For example, the ratio of rest-frame 24-$\mu$m to radio
luminosity, $q_{24}$, is often used to discriminate between radio-loud
and radio-quiet sources \cite{ibar08}, but the 9.7-$\mu$m silicate
absorption feature passes through the 24-$\mu$m filter bandpass at
$z \sim 1.5$, making the $k$-correction very sensitive to the adopted
redshift and possibly causing the misclassification of sources.

A final, yet important, concern regarding a comprehensive analysis of
the faint radio source counts arises from a lack of uniformity in the
measurement of source fluxes. The scatter observed between the
different measurements of the source counts in
Fig.~\ref{fig:sourcecounts} is much larger than can be attributed to
cosmic variance \cite{dezotti10,heywood13}.  Most source flux
densities are determined (at least as a first estimate) from an
elliptical Gaussian fit, where there exist correlations between the
fitted parameters. For a genuinely unresolved source, these fits
overestimate the true flux density if the best-fitting source size is
larger than the beam, and underestimate it if the source is smaller
than the beam \cite{condon97}. However, some authors ignore the fit
when it produces a source size smaller than the beam (which is
obviously unrealistic) and adopt the peak value as the flux density,
but this overestimates the true flux density. Consequently, the fluxes
of \textit{all\/} unresolved sources are overestimated, except in the
very unlikely case that the fitted source size is exactly the beam
size. With the source counts being so steep below 1\,mJy, systematic
overestimates of source fluxes are amplified when used to derive the
source counts: since the counts are roughly flat when plotted in the
manner of Fig.~\ref{fig:sourcecounts}, a 10\% overestimate of the
source fluxes will result in a 27\% overestimate of the source counts,
and even larger flux errors are possible (see fig.~4 of
ref.~\cite{cjs06}). Furthermore, resolved sources can be missed in a
catalogue if their peak flux densities are below the detection
threshold, and correcting for this incompleteness requires knowledge
of the true source size distribution, which remains rather uncertain
\cite{windhorst90,bondi03}. Recovering integrated flux densities from
a tapered lower-resolution map is impractical in the deepest surveys
due to confusion, and it is vital that agreement is reached on how to
measure reliable source fluxes from interferometric data, or the
scientific usefulness of these data will be severely compromised.

\section{The radio-loud/radio-quiet distinction}
\label{sec:rlrq}

One of the major questions in extragalactic astronomy is the origin of
radio-loudness. Although the first quasars were discovered because of
their powerful radio emission, most optically-selected quasars were
too faint to be detected at radio wavelengths. Once the VLA enabled
sensitive observations of large samples of optically-selected quasars,
a dichotomy in their radio properties appeared. Defining
radio-loudness, $R$ as the ratio of radio to optical luminosities
(typically $B$-band and 5\,GHz), there appeared to be a dearth of
sources with $R \approx 1$--10
\cite{marshall87,kellermann89}\footnote{Some authors have tried to use
  this ratio as a diagnostic for non-quasar sources but the optical
  emission of these objects arises from the stellar population, whose
  luminosity does not generally correlate with accretion power
  (although for Eddington-limited sources, the two are connected via
  the central black hole mass).}. Sources with large values of $R$
were determined to be an orientationally-biased subset of powerful
FR\,II radio galaxies, with their true radio structures often obscured
by the powerful Doppler boosting of the core.

Radio luminosity is easy to measure but is known to be only a small
fraction of the energy needed to power the jets in these powerful
sources. Two methods for estimating the kinetic power of the
large-scale radio jets have been widely used in the literature. One
uses the minimum energy condition to estimate the energy stored in the
particles and magnetic field \cite{miley80}, and then divides this by
the estimated age of the radio source \cite{rawlingssaunders91}. A
second, more recently-developed, method is applicable in cases where
the radio source has excavated bubbles in the surrounding
X-ray-emitting plasma. Here, the $p\,dV$ work done in excavating the
bubbles is estimated and, again, is divided by the estimated age of
the source \cite{birzan04,cavagnolo10}. Given the observed scatter and
theoretical uncertainties in these relationships, the precise
normalization of the relationship between radio luminosity and kinetic
jet power is uncertain by a factor of a few but, in general, results
support a relationship $Q \propto L_\nu^\beta$ where $\beta \approx
0.8$ \cite{heckmanbest14}.  Estimates for powerful radio-loud sources
suggest that the efficiency for converting kinetic power, $Q$, to
radio luminosity, $\nu L_\nu$, is of the order of one per cent,
assuming that the jets contain only light particles \cite{bicknell95},
and this power is comparable to the total luminosity in ionizing
photons \cite{willott99}.

The ability to use radio luminosity as a proxy for the accretion power
in powerful radio sources, albeit with significant normalization
uncertainties, is highly beneficial since the radio photons are
unaffected by intervening dust or gas, but it remains unclear whether
there is a similar relationship for radio-quiet objects. In radio-loud
sources, the radio emission unequivocally arises from
synchrotron-emitting plasma that is transported by large-scale
jets. The radio emission in radio-quiet objects is much smaller in
extent and approximately three orders of magnitude lower in
luminosity, when normalized by an accretion-rate-dependent quantity
such as optical luminosity (for quasars) or narrow-line luminosity.
This naively suggests that radio-quiet sources have an energy input
into synchrotron-emitting particles $\sim 1000$ times lower, but that
assumes the physics are the same and it is unclear whether the radio
emission in these objects arises from plasma accelerated in jets.
There have been two recent suggestions for non-jet origins of the
radio emission in radio-quiet objects. Similarities between the shapes
of the RLFs for nearby galaxies and low-redshift QSOs were used to
suggest that the radio emission in QSOs might arise from star
formation \cite{kimball11}. Such objects would only need modest star
formation rates of $\sim20$\,M$_\odot$\,yr$^{-1}$ but distant luminous
QSOs would require rates of hundreds of M$_\odot$\,yr$^{-1}$
\cite{condon13}. With mid-infrared observations apparently ruling this
out, it has instead been suggested that the radio emission in
radio-quiet sources may arise from wind-driven shocks, which would be
expected to have a very low radiative efficiency by analogy with
supernova-driven shocks \cite{zakamska16}. While radio-quiet quasars
display a range of morphologies, some display very clear jet-like
features and appear to simply be scaled-down versions of their
radio-loud cousins \cite{miller93,kellermann94}. In nearby Seyfert
galaxies, where the physical resolution is better, linear structures
are seen in three-quarters of objects whose radio emission is resolved
\cite{ulvestad84}. There is also a tendency for linear structures to
be more prevalent in more radio-luminous objects which could arise
either as a consequence of the correlation between linear size and
radio luminosity coupled with limited angular resolution, or simply
because lower-luminosity jet-like features can be more easily swamped
by the radio emission from the star formation that is often associated
with AGN activity.

Perhaps the most compelling evidence for jets in radio-quiet QSOs
comes from VLBI observations that reveal extremely high brightness
temperature cores \cite{kmbbeasley98,herreraruiz16}, while the
abundance and properties of optically-selected flat-spectrum QSOs are
most readily explained if the radio emission arises from Doppler
boosting, requiring the jets to be relativistic
\cite{falcke96a,falcke96b}. The quasar E1821+643 is a particularly
interesting source because it has such a high optical luminosity ($M_B
\approx -27$) that, despite its radio quiet nature, its radio
luminosity is similar to that of nearby FR\,I radio galaxies, and it
displays a similar radio morphology \cite{e1821mdl,e1821kmb}. This
suggests that perhaps radio-quiet active galaxies simply supply a much
smaller fraction of their accretion energy to producing jets, which
are otherwise identical to those of radio-loud objects. Alternatively,
the particular make-up of the jets might differ, with `light'
electron--positron jets being launched in radio-loud sources and
`heavy' electron--proton jets launched in radio-quiet ones. The kinetic
jet powers of both types of source would therefore be similar, but
$\sim 99.9$\,per cent of the kinetic energy in radio-quiet objects
would be used to accelerate protons that emit negligible synchrotron
emission.

Could the difference in the synchrotron luminosities of radio-loud and
radio-quiet objects be driven by a difference in the properties of the
supermassive black holes that power them? It is known that radio-loud
objects are limited to the sources with the highest accretion rates,
as identified by [O{\sc~iii}] emission \cite{miller93}, and the
highest black hole masses \cite{laor00,rjmmjj04}, but these extreme
objects are not \textit{exclusively\/} radio-loud. It has also been
known for many years that the hosts of powerful radio-loud objects are
exclusively elliptical galaxies and these facts have been brought
together by a number of authors to suggest that the energy source for
powerful radio jets may be the rotational angular momentum of the
black hole \cite{bz77,wilson95,sikora07,amssr11}. If the primary route
for creating rapidly-spinning black holes is via the merger of two
similar-mass black holes (presumably hosted in two similar-mass
galaxies) where orbital angular momentum is converted to rotational
angular momentum during coalescence, then it follows that such objects
will reside in ellipticals (the product of major galaxy mergers) and
will typically have higher masses than the black holes that power
radio-quiet sources.

While this is a qualitatively attractive model, it suffers from the
lack of a quantitative formalism for the energetics of jet production.
In addition, indirect measurements of black hole spin in nearby
Seyfert galaxies are high despite the radio-quiet nature of these
objects \cite{risaliti13,agisgonzalez14,wang17}, although doubt has
been cast on the validity of these results \cite{millerturner13}. An
alternative idea suggests the radio-loud and radio-quiet states are a
form of duty cycle analogous to those seen in X-ray binaries
\cite{maccarone03,nipoti05,kording06}. An apparent prediction of this
scenario is that AGN should be able to shut off and then restart in a
different mode, in particular to transition from radio-loud to
radio-quiet. While there are known to be sources that stop and restart
their activity while remaining radio-loud (so-called `double-double
radio galaxies'), no sources are known where a radio-quiet AGN is
surrounded by relic lobes from a prior radio-loud episode of activity.

\section{Distant radio sources}

There already exist a number of excellent reviews of high-redshift
radio galaxies (HzRGs) \cite{mccarthy93,mileydebreuck08}. Known
examples of these objects are exclusively luminous high-excitation
FR\,II sources whose bright emission lines (including Ly$\alpha$)
allow for easy redshift measurements. Apart from the general desire of
astronomers to hold the record for discovering the ``most \ldots''
something, there are compelling scientific reasons for searching for
HzRGs: the evolution of the radio galaxy population at early cosmic
times will reveal details about the formation mechanism of these rare
objects; being powered by the most massive supermassive black holes,
they are likely to form at density peaks and provide signposts to
galaxy protoclusters; and sufficiently distant objects ($z > 6$) can
act as background radiation sources against which to study
reionization via the 21-cm forest.

Given the extreme rarity of such objects among the general radio
source population, various criteria have been developed to improve the
efficiency of HzRG searches by removing objects likely to be in the
foreground. Typically sources larger than $\sim 15''$ are excluded,
due to an observed anti-correlation between redshift and angular size
\cite{kapahi89} and, since distant radio galaxies are also found to
have steeper spectral indices \cite{blumenthal79}, a spectral-index
cut is also applied. Following up only sources with $\alpha > 1$
enabled the identification of the first $z>2$ radio galaxies
\cite{chambers87,chambers88} and even more aggressive spectral index
cuts produce a higher fraction of the highest redshift sources
($z>3$), with the extreme criterion of $\alpha > 1.3$ enabling the
discovery of the most distant radio galaxy identified to date,
TN~J0924--2201 at $z=5.19$ \cite{tnj0924,debreuck01}. Of course, such
aggressively-selected samples suffer from hard-to-quantify
incompleteness and cannot be reliably used to estimate the space
density of HzRGs \cite{mjj01}.

By modelling the evolution of powerful radio sources, the reason for
these correlations with redshift was traced to dramatically increased
inverse Compton losses to the Cosmic Microwave Background (the energy
density of the CMB increases $\propto (1+z)^4$)
\cite{kaiser97,kmbsr99,kmb99}. Since more energetic electrons (which
emit their synchrotron radiation at higher frequencies) lose energy to
inverse Compton scattering more rapidly, this results in the radio
spectrum becoming steeper and more curved, as well as decreasing in
luminosity. Since a single observing frequency samples increasing
rest-frame frequencies for higher redshift objects, the HzRG fraction
will be largest in low-frequency selected catalogues.

Even with these selection criteria, a sample optimized for finding
HzRGs based solely on radio properties will still contain many
lower-redshift interlopers. Only optical or near-infrared imaging can
reliably exclude these objects, as true HzRGs will be very faint at
these wavelengths. $K$-band imaging is most frequently used since
radio galaxies have a tight locus in the near-infrared Hubble diagram
and the so-called `$K$--$z$ relation' \cite{willott03} can be used to
give a very crude redshift estimate, often providing support for the
identification of a single emission line as Ly$\alpha$ rather than
[O{\sc~ii}] at a much lower redshift.

The Hubble diagram for radio galaxies was originally developed to
measure the Hubble constant, under the assumption that the sources
were all first-ranked ellipticals and could be used as standard
candles \cite{sandage72i,sandage72iii}. Historically, therefore, it
was parametrized as $m(z)$, with the magnitude measured in an aperture
of fixed linear size of 63.9\,kpc, corresponding to approximately
$8''$ at high redshift. Given the faintness of distant radio
galaxies, the unacceptable loss of signal-to-noise ratio incurred from
the use of such a large aperture requires authors to make measurements
in smaller apertures and correct these to the standard aperture using
a curve of growth. Since distant radio galaxies clearly do not follow
this smooth surface brightness profile, the results are dependent on
the actual aperture used, as well as the adopted curve of growth and
cosmology. Progress has been slow in moving towards a more practical
Hubble diagram, using magnitudes in a fixed angular aperture
\cite{bryant09}, as well as fitting the redshift as a function of
magnitude \cite{brookes08}, which is more appropriate since the
diagram is now used to estimate redshifts, with the observed magnitude
being the independent variable.

Obtaining the deep imaging needed to exclude low-redshift sources is
the most expensive step (in terms of telescope time) in the
identification process. With typically an hour of 8-m telescope time
required per source, very severe size and spectral index cuts are
needed to filter a sample to provide high efficiency, at the expense
of completeness. Rather than pursue bespoke imaging for filtered
samples of radio sources, however, it is possible to use existing deep
near-infrared imaging with less severe radio selection, allowing a
reliable measurement of the space density of radio galaxies in the
early Universe. Cross-correlating the \textit{Spitzer Space
  Telescope\/}'s SWIRE survey \cite{swire} with the FIRST \cite{first}
survey produced a sample of infrared-faint radio galaxies over
24\,deg$^2$ with counterparts ready for spectroscopy. Although the
1.4-GHz frequency of the FIRST survey is not optimal for finding
HzRGs, three new $z>4$ radio galaxies were identified
\cite{teimourian}, including the second most distant radio galaxy
known \cite{jarvis09}. This source, FIRST~J163912.11+405236.5 at
$z=4.88$, has a spectral index between 325\,MHz and 1.4\,GHz of 0.75,
which would have excluded it from any sample filtered by spectral
index. A larger sample of objects selected with similar criteria
further demonstrated these objects do not typically have ultra-steep
radio spectra \cite{collier14}, although the very faintest infrared
sources typically do have $\alpha > 1.0$, suggesting they may lie at
the very highest redshifts \cite{middelberg11}.

Although the record for the most distant radio source has stood for
nearly two decades, models predict $\sim 200$ radio galaxies with $z
>6$ and $S_{150} > 10$\,mJy over the sky. Such sources can provide a
unique insight into the Epoch of Reionization (EoR) by observing the
21-cm forest in absorption \cite{carilli02}. This is analogous to the
use of the Ly$\alpha$ forest in absorption the lines-of-sight to
optically-bright quasars \cite{fan06} but is much less limited.
Whereas a neutral fraction as low as $\sim$0.01\,per cent leads to an
opaque Gunn--Peterson trough within which even the deepest
spectroscopic observations cannot detect any flux \cite{barnett17},
the 21-cm hyperfine transition has a cross-section $\sim10^7$ times
smaller than Ly$\alpha$, and even predominantly neutral gas transmits
radiation, allowing the clumpiness of the intergalactic medium to be
studied. Optical/near-infrared searches for high-redshift quasars
continue to discover new sources at $z>6$ but these are all
radio-quiet and therefore too faint for 21-cm forest studies. It is to
be hoped that, as the number of such sources increases towards $\sim
100$, one or more is found to be radio-loud simply by chance.

At less extreme redshifts, the apparent tendency of radio galaxies to
live in dense environments and hence act as signposts to protoclusters
provides another reason to search for them. The first observational
clues that at least some HzRGs lived in dense environments came from
the large rotation measures observed \cite{carilli97}. Deep imaging
often reveals an excess of companion sources, identified via
photometric redshifts, narrow-band imaging for Ly$\alpha$ or H$\alpha$
emission, or spectroscopy. While a significant number of HzRGs are
known to reside in protoclusters, including TN~J0924--2201
\cite{venemans04,overzier06} and the well-studied `Spiderweb' galaxy
PKS~1138--262 \cite{kurk00,kuiper11}, there is no measurement of the
fraction of HzRGs that live in such environments from an unbiased
sample. Furthermore, since the lifetime of the radio source
($\sim10^8$\,yr) is much less than the cluster collapse time (a few
Gyr), the epoch of radio source activity may highlight a specific
period in the formation of the cluster. While these distant
protoclusters therefore provide interesting laboratories in the early
Universe, one must be cautious of possible biases that might preclude
generalizing any results.

\section{The future}

Three new radio telescopes are in the process of conducting major
continuum surveys with equivalent depths $S_{1400} \ll 1$\,mJy over
large regions of the sky. These surveys will be dominated by
star-forming galaxies and a key requirement in extracting the maximum
scientific return will be identifying and separating these objects
from other classes such as radio-quiet AGN. Although high-resolution
radio observations ($\sim 0.1''$, requiring 500\,km baselines at
1.4\,GHz) can reveal the presence of an AGN that eludes detection at
all other wavelengths \cite{casey09}, the surveys have resolutions of
a few arcseconds, preventing the use of radio morphology as a
discriminant, and the radio spectra of both classes of source are
similar. Spectroscopy will be useful in assigning objects to the
``mainly star-forming'' or ``mainly AGN'' bins but this is still prone
to bias from dust obscuration. Given the vast sizes of the radio
source samples created by the pathfinders, follow-up is likely to be
either statistical, with subsamples used as training sets for machine
learning, or focused on unusual objects that lie outside the normal
parameter space.

The Australian SKA Pathfinder,
ASKAP\footnote{http://www.atnf.csiro.au/projects/askap}, consists of
36 12-m antennas equipped with phased array feeds that dramatically
increases its survey speed. Among several Survey Science Projects, the
Evolutionary Map of the Universe survey (EMU \cite{emu}) is an
`all-sky' (3$\pi$ steradians) survey at 1.3-GHz with 10$''$ resolution
and 10\,$\mu$Jy rms. The reliability of identifying the
optical/infrared counterparts to radio sources does not suffer
significantly at this resolution, but the depth of the complementary
data has a major effect \cite{mcalpine12}. This will be the main
limitation in trying to extract the maximum from all-sky surveys and,
while the Large Synoptic Survey Telescope (LSST) will ultimately
survey the entire southern sky to $r\sim27.5$, similar to the SXDS
optical data, these observations will not be completed until the
2030s.

The African SKA precursor telescope,
MeerKAT\footnote{http://www.ska.ac.za}, will comprise 64 13.5-m
antennas. These are fitted with single-pixel detectors, making the
telescope better suited to deep surveys over small areas. Most
relevant to this article is the MIGHTEE survey, which will cover
20\,deg$^2$ at 1.4-GHz to a depth of 1\,$\mu$Jy rms in fields where
there is the best complementary data, including the LSST deep-drilling
fields and those with near-infrared imaging from the VIDEO survey
\cite{video}, to aid with the cross-identifications.

The Low Frequency Array (LOFAR\footnote{http://www.lofar.org}) is
already conducting surveys between 20 and 200\,MHz, with its most
sensitive observations around 150\,MHz. It is undertaking several
nested surveys at multiple frequencies, from a shallow `all-sky'
survey to a deep survey covering a few individual pointings centered
on fields with excellent complementary data. It also has a dedicated
spectroscopic follow-up programme, the WEAVE/LOFAR survey
\cite{weavelofar}, which will produce approximately one million
spectroscopic redshifts for radio sources selected at 150\,MHz. The
low-frequency selection favours the identification of the highest
redshift sources and the widest tier of the WEAVE/LOFAR survey is
expected to find tens of radio galaxies at $z>6$, producing multiple
lines of sight to probe the structure of the intergalactic medium
during the Epoch of Reionization using the 21\,cm forest.

In addition to these new facilities, the Karl G.\ Jansky Very Large
Array is undertaking the VLA Sky Survey
(VLASS\footnote{http://science.nrao.edu/science/surveys/vlass}), a
3-GHz survey covering 82\,per cent of the celestial sphere in two
epochs with a resolution of 3$''$ and a depth of 80\,$\mu$Jy rms.
These four projects together provide complementary new windows on the
radio sky from their sky area, depth, and/or observing frequency, and
will help to define the optimal SKA surveys as well as answering
scientific questions themselves.

Over the past decade, there has already been a gradual shift in the
use of radio data, as its synergy with imaging at other wavelengths
has been exploited. As we approach and enter the era of the SKA, the
properties of the radio sky should become essential information for
all extragalactic astronomers. The density of radio sources on the sky
with $S_{1400} > 1\,\mu$Jy is comparable to the number of galaxies
with $V < 24$ so radio fluxes will be available for the majority of
objects detected in optical/near-infrared surveys and represent
important photometric data, providing a dust-independent measurement
of the star-formation rate, or indicating the presence of an AGN.
Modelling the radio sky in simulations and semi-analytic models should
also be a goal, but this requires an understanding of the physical
processes that trigger the radio emission from galaxies, and this is
still highly uncertain except for star-formation. Since supermassive
black holes are now included in simulations and semi-analytic models,
there is a possibility that, as these models mature, they could help
to reveal the cause of the radio-loud/radio-quiet dichotomy in AGN.

As much of a challenge as the raw data processing requirements of
these new instruments will be learning how to use the vast array of
information efficiently to answer questions, some of which have not
even been posed yet. Radio astronomy has often been seen as an obscure
and separate discipline but the next few years represent a key period
in integrating the subject into the mainstream. The Square Kilometre
Array will be a truly global telescope and its data should be owned by
everybody.

\section*{Competing Interests}

I have no competing interests.

\section*{Funding}

No funding was provided for this work, although it was carried out
with the support of the Gemini Observatory, which is operated by the
Association of Universities for Research in Astronomy, Inc., on behalf
of the international Gemini partnership of Argentina, Brazil, Canada,
Chile, and the United States of America.

\section*{Acknowledgments}

\ack{I would like to thank Philip Best, Katherine Blundell, and Matt
  Jarvis for commenting on the first draft of this manuscript. I am
  also grateful for the support I have received in recent times from
  friends and colleagues who deal in facts and evidence, in particular
  Dan Smith and Eve and James Ewington, without whom\ldots\ Finally,
  thanks to Rob Ivison for giving me the opportunity to write this
  review, and to Steve Rawlings, for turning the dome.}



\end{document}